\documentclass[twocolumn,iop,numberedappendix,appendixfloats]{openjournal}
\submitted{}

\makeatletter
\renewcommand{\frontmatter@title@below}{\vspace*{-2.63\baselineskip}\vspace*{0.65in}}
\makeatother

\UseRawInputEncoding
\usepackage{xcolor}
\usepackage{textgreek}
\usepackage[utf8]{inputenc}
\usepackage[english]{babel}
\usepackage{booktabs}
\usepackage{comment}
\usepackage{orcidlink}
\usepackage{bm}

\usepackage{hyperref}
\hypersetup{
    unicode, 
    colorlinks=true,
    linkcolor=linkcolor,
    citecolor=linkcolor,
    filecolor=linkcolor,
    urlcolor=linkcolor,
}
\usepackage{color,colortbl}
\definecolor{linkcolor}{rgb}{0.0,0.3,0.5}
\usepackage{tensind}
\tensordelimiter{?}
\DeclareGraphicsExtensions{.bmp,.png,.jpg,.pdf}
\usepackage{amsmath}

\newcommand{\Euclid}{\emph{Euclid}}
\newcommand{\Merlin}{\texttt{Merlin}}
\newcommand{\Nautilus}{\texttt{Nautilus}}
\newcommand{\Swyft}{\texttt{Swyft}}
\newcommand{\cloelib}{\texttt{cloelib}}
\newcommand{\threetwopt}{3$\times$2pt}
\newcommand{\thet}{{\bm{\theta}}}
\newcommand{\et}{{\bm{\eta}}}
\newcommand{\bs}{{\bm{S}}}
\newcommand{\varthet}{{\bm{\vartheta}}}
\newcommand{\xx}{{\bm{x}}}
\newcommand{\lcdm}{$\Lambda$CDM}

\graphicspath{ {./figs/} }

\begin{document}
\title{MERLIN: Fast and flexible \threetwopt~cosmology with simulation-based inference}

\author{Alexandra Wernersson}
\email[*]{alewer97@gmail.com}
\affiliation{Nikhef, Science Park 105, 1098 XG, Amsterdam, The Netherlands}
\affiliation{CAISR Health, Halmstad University, Halmstad 30118, Sweden}

\author{Guillermo Franco-Abell\'an\orcidlink{0000-0001-9742-5408}}
\email[$\dag$]{g.francoabellan@ific.uv.es}
\affiliation{IFIC, CSIC-Universitat de Val\`encia, c/ Catedrático José Beltrán, 2 E-46980 Paterna, Spain}

\author{Guadalupe Cañas-Herrera\orcidlink{0000-0003-2796-2149}}
\email[$\star$]{canasherrera@strw.leidenuniv.nl}
\affiliation{Leiden Observatory, Leiden University, PO Box 9506, Leiden 2300 RA, The Netherlands}

\begin{abstract}
We present \Merlin, a simulation-based inference (SBI) pipeline to perform cosmological analyses of \threetwopt~summary statistics: cosmic shear, galaxy clustering and the cross-correlation power spectra. Our approach combines Marginal Neural Ratio Estimation (MNRE) with \threetwopt~angular power spectra predictions from the \cloelib~library, although the pipeline is readily extensible to other cosmology libraries. We demonstrate this pipeline on a realistic setting representative of a Stage-IV photometric survey, with a 50-dimensional parameter space describing cosmology and a wide range of systematic effects. We find posteriors that are in excellent agreement with the nested sampler \Nautilus, while reducing the number of required CPU-hours by two orders of magnitude. Because the generation of training data and the training of inference networks are decoupled processes, a single simulation bank can be reused to perform inference under different analysis choices, further improving the simulator-efficiency. We illustrate this flexibility by applying scale cuts, varying the survey area, and removing cosmic shear from the data vector (2$\times$2pt), all at \emph{zero} extra model evaluations, whereas sampling-based methods need costly re-runs for each case. The \Merlin~code is publicly available on GitHub\footnote{\url{https://github.com/Alexandra-Wernersson/merlin}}. 
\end{abstract}


\maketitle

\section{Introduction}
\label{sec:intro}

The current Stage-IV generation of wide-field photometric surveys --- including ESA's \Euclid~space mission \citep{EUCLID:2011zbd, Euclid:2024yrr}, the Vera C.\ Rubin Observatory Legacy Survey of Space and Time \citep{LSST:2008ijt} and the NASA's Nancy Grace Roman space telescope --- will image billions of galaxies over thousands of square degrees, delivering unprecedented statistical power for precision cosmology.  A cornerstone of these analyses is the joint exploitation of three two-point summary statistics (\threetwopt): weak gravitational lensing (WL, also known as cosmic shear), photometric galaxy clustering (GCph), and the cross-correlation, dubbed galaxy--galaxy lensing (GGL). Analysed together, this \threetwopt~data vector simultaneously self-calibrates astrophysical nuisance parameters and improves the constraining power on cosmology, with extra impact on the amplitude of matter fluctuations $\sigma_8$, the total matter density $\Omega_{\rm m}$, and the dark-energy equation-of-state parameters $w_0$ and $w_a$ \citep{DES:2021wwk, Wright:2025xka, DESI:2024mwx, DES:2026fyc, DES:2026jmi, Euclid:2025pzh}. 

Classical Bayesian inference methods for \threetwopt~conventionally proceed by evaluating the likelihood of the data given the model parameters, which is is typically assumed to be a Gaussian with a covariance matrix fixed at a fiducial cosmology. This framework has proved remarkably successful in Stage-III surveys, but it faces two interconnected challenges as data quality improves.  

First, it has been shown that the likelihood for cosmic shear power spectra can be substantially different from the Gaussian that is commonly assumed \citep{Oehl:2026xgc}. Even in the cases where the Gaussian likelihood approximation is justified, its evaluation  requires the output of a Boltzmann solver together with some recipe to model non-linear scales, which is time-consuming (specially for extended cosmologies).  Second, the parameter space of a realistic \threetwopt~analysis in Stage-IV surveys is very high-dimensional: beyond the few cosmological parameters of interest, one must simultaneously marginalize over intrinsic alignments, galaxy bias, magnification bias, multiplicative shear-calibration uncertainties, and photometric redshift systematics, collectively amounting to $\mathcal{O}(30\text{--}65)$ nuisance parameters. Exploring such a space with traditional Markov Chain Monte Carlo (MCMC) methods \citep{Foreman-Mackey:2012any} or nested samplers \citep{Feroz:2008xx, Handley:2015fda, Speagle:2019ivv, Lange:2023ydq} requires $\mathcal{O}(10^6\text{--}10^7)$ likelihood evaluations, making the total inference cost computationally prohibitive. Furthermore, many extended cosmological models introduce complex parameter degeneracies and highly non-Gaussian posteriors, which further slows down the convergence. 

One way to address these challenges is to use trained emulators, which enable extremely fast likelihood evaluations by emulating the matter power spectrum \citep{Euclid:2020rfv,Piras:2023aub,Nygaard:2024lna,Gunther:2025xrq} or directly the  two-point angular statistics \citep{Bonici:2022xlo}. Inference can be further optimized if emulators are combined with MCMC methods that are scalable to high-dimensional parameter spaces, like Hamiltonian Monte Carlo (HMC) \citep{Piras:2024dml,Nygaard:2026fgl, Bonici:2025ltp}. These methods achieve a highly efficient exploration of the parameter space by leveraging  the information coming from the gradient of the likelihood with respect to the parameters. However, this strategy retains some of the caveats inherent to any likelihood-based method, like the need to assume a potentially incorrect functional form of the likelihood (which must be differentiable for HMC).

Alternatively, modern approaches in simulation-based inference (SBI) offer a principled route around these data analyses challenges without requiring an explicit evaluation of the likelihood (see \citealt{Cranmer2020} for a review).  Instead of a likelihood, the key input in SBI is a simulator that maps from model parameters $\thet$ to data realizations $\xx$; these data--parameter pairs $(\thet,\xx)$ are typically used as training data for neural networks that learn either  the posterior density \citep{Papamakarios2019,Jeffrey:2020itg}, the likelihood \citep{Papamakarios2016}, or the posterior-to-prior ratio \citep{Cranmer:2015bka,Hermans2020}. Hence, SBI makes inference possible even when the likelihood cannot be expressed analytically or it is very hard to compute. SBI brings some extra key advantages: i) it can directly target the marginal posteriors of interest, without having to sample the full joint posterior, which makes it scalable to high-dimensional problems; ii) the generation of simulations can be fully parallelised and hence made extremely fast given appropriate resources; and iii) it can enable amortized inference, meaning that the trained posterior applies to any observation drawn from the prior, allowing for rapid re-analysis and cross-validation.

Applications of SBI to cosmological analysis of large-scale structure (LSS) probes have grown rapidly over the past few years. In particular, SBI has been successfully applied to weak-lensing maps \citep{DES:2023qwe,DES:2024xij,Novaes:2024dyh,vonWietersheim-Kramsta:2024cks,Zeghal:2024kic,DES:2026vyi}, galaxy clustering \citep{Hahn:2022zxa,SimBIG:2023ywd,Modi:2023drt,Tucci:2023bag,Hou:2024blc}, and combined probes in simulated survey settings \citep{DES:2025jhz}.  However, a complete SBI pipeline for realistic \threetwopt~analyses of a Stage-IV galaxy survey like \Euclid, encompassing marginalisation over a large number of nuisance parameters and a direct comparison with state-of-the-art samplers, has not yet been demonstrated.

In this paper we introduce \Merlin, a modular, end-to-end SBI pipeline for fast and flexible cosmological analysis of \threetwopt~statistics in realistic Stage-IV settings. \Merlin~is built on top of \Swyft~\citep{Miller2021} to perform Marginal Neural Ratio Estimation (MNRE), and, currently, it couples with the \cloelib~library\footnote{\href{https://github.com/cloe-org/cloelib}{https://github.com/cloe-org/cloelib}} \citep{Euclid:2026dlz} to produce angular power spectra predictions used to train the neural networks. To analyze the hundreds of power spectra that comprise the data vector, the code uses the same pre-compression strategy that was developed in \cite{FrancoAbellan:2024tbj}. 

We note that contrary to most works applying SBI to LSS, which have been focused on intractable likelihoods (e.g. field-level analyses or the modeling of complex systematics), the simulator in \Merlin~samples data vectors from a multivariate Gaussian with given covariance, making it equivalent to an explicit Gaussian likelihood. At first glance, this choice might appear somewhat peculiar, but there are two important reasons behind it: i) it allows for a direct comparison with standard samplers, which is a crucial step before moving to a more realistic simulator capturing any non-Gaussianities in the likelihood; and ii) SBI brings strong advantages even for explicit likelihoods. To illustrate this, we show that our approach can efficiently marginalize over a 50-dimensional (5 cosmological and 45 nuisance) parameter space and yield accurate posteriors in a fraction of the time needed by the widely used \Nautilus~nested sampler \citep{Lange:2023ydq}. Moreover, we highlight a key strength of SBI that has often been overlooked, namely the fact that simulations can be reused across inference tasks, which eliminates the need for repeated model evaluations \citep{FrancoAbellan:2025fkb}. We show that the same simulation bank can be recycled to consider multiple variants in the analysis (scale cuts, different survey area, removing WL from the data vector) while requiring \emph{zero} additional model evaluations.

The paper is structured as follows.  In Sect.~\ref{sec:cells} we discuss the theoretical modeling of the \threetwopt~summary statistics, and in Sect.~\ref{sec:sbi} we describe in detail the components of the \Merlin~inference pipeline. Our main results are presented in Sect.~\ref{sec:results}, including posterior comparisons with \Nautilus~and a discussion of computational costs.  We draw our conclusions and outline future developments in Sect.~\ref{sec:concl}.

\begin{figure*}[ht!]
	\centering
	\includegraphics[width=\textwidth]{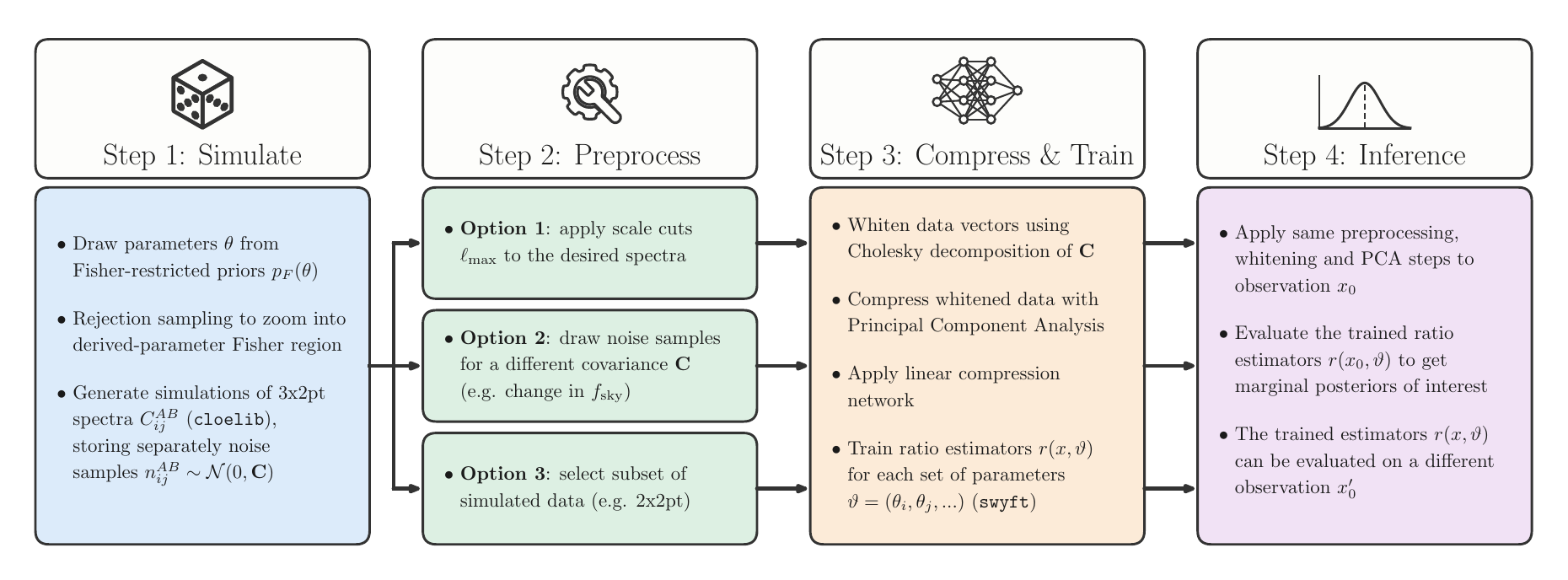}
	\caption{Overview of the \Merlin~pipeline. In Step 1, simulated \threetwopt~spectra are generated by sampling from a Fisher-truncated prior and calling \cloelib, together with noise samples drawn from a given covariance matrix. These simulations can be optionally preprocessed in Step 2 to explore variations in the analysis (e.g. scale cuts $\ell_{\rm max}$, a different covariance matrix, or 2$\times$2pt only). In Step 3, we jointly train networks to compress the data and perform Marginal Neural Ratio Estimation via the \Swyft~code. Finally, in Step 4 the trained networks are evaluated on a given observation to obtain posterior constraints.}
	\label{fig:pipeline}
\end{figure*}

\section{Data vector}
\label{sec:cells}

The \threetwopt~harmonic space data vector is formed by concatenating the angular power spectra of three pairs of tracers evaluated at a common set of multipoles $\ell$:
\begin{equation}
    \xx = \bigl[C^{\rm WL}_{ij} (\ell),\; C^{\rm GGL}_{ij} (\ell),\; C^{\rm GCph}_{ij} (\ell)\bigr],
\end{equation}
where $i,j=1,...,N_{\mathrm{bin},z}$ run over tomographic redshift bins. We work with $N_{\mathrm{bin},z} = 13$, as suggested in \citet{Euclid:2024yrr} for a Stage IV photometric survey configuration by the end of operations.  The WL auto-spectra comprise all $N_{\mathrm{bin},z}(N_{\mathrm{bin},z}+1)/2$ auto- and cross-bin combinations of the shear field, and the same holds for GCph.  The GGL cross-spectra include all $N_{\mathrm{bin},z}^2$ combinations of the galaxy-position and shear tracers. Hence, this results in a total of 351 independent angular power spectra, each of which is evaluated in $32$ angular bins.

Angular power spectra are computed with the \cloelib~\texttt{AngularTwoPoint} module using the Limber approximation. To accelerate calculations, perturbations are evaluated with the \texttt{HMcode2020} emulator \citep{Mead:2020vgs, Tsedrik:2024cdi}, which provides the non-linear matter power spectrum $P_{\rm nl}(k,z)$ as a function
of the cosmological parameters and the baryonic feedback parameter $\log_{10}T_{\rm AGN}$. Background quantities for shear and position kernels are computed with the Boltzmann solver \texttt{CAMB} \citep{Lewis:1999bs, Lewis:2026mif}.

The shear tracer includes the non-linear intrinsic alignment (NLA) model with amplitude $A_{\rm IA}$ and redshift-evolution index $\eta_{\rm IA}$ \citep{Bridle:2007ft}, multiplicative shear-calibration biases $m_i$ for each tomographic bin $i$, and photometric-redshift shift parameters $\Delta z_i$.

The galaxy-position tracer includes a polynomial galaxy-bias model with four free coefficients $\{b_{g1},\ldots,b_{g4}\}$, per-bin magnification-bias parameters $\{b_{{\rm mag},i}\}$, and per-bin photometric-redshift shifts $\Delta z_i$. We assume that the photometric-redshift shift of a given tomographic bin is shared between the
position and shear tracers, $\Delta z_i \equiv \Delta z_{{\rm pos},i} = \Delta z_{{\rm she},i}$.  For $N_{\rm bin}=13$ tomographic bins the full nuisance parameter vector therefore has dimension $2 + 4 + 3 N_{\rm bin} = 45$, comprising the two intrinsic alignment parameters, the four galaxy-bias coefficients, and the three per-bin systematics (magnification bias, multiplicative shear calibration, photometric-redshift shift).

For this analysis, we assume a spatially flat $\Lambda$CDM cosmology, with the fiducial cosmological parameters and systematic effects specified in Tab.~\ref{tab:priors}. The prior distributions are chosen to reflect the expected survey calibration precision and are summarised in the same table. Cosmological and galaxy-bias parameters are assigned uniform priors, while the multiplicative shear-calibration parameters $m_i$ are assigned zero-mean Gaussian priors with $\sigma_m = 0.01$. Photometric-redshift shifts $\Delta z_i$ are likewise assigned Gaussian priors with $\sigma_{\Delta z} = 0.002(1+\bar{z}_i)$, where $\bar{z}_i$ denotes the mean redshift of bin $i$, with values $[0.289, 0.376, 0.437, 0.536, 0.619, 0.709, 0.802, 0.859, 0.976$, $1.093, 1.246, 1.489, 1.922]$. This reference model is inspired on the specifications of \citet{Euclid:2024yrr} and \citet{Euclid:2025pzh}. We adopt the same prior distributions for both \Merlin~and \Nautilus~to enable an apples-to-apples comparison. However, as it will be discussed in Sect.~\ref{sec:fisher}, the priors for \Merlin~are restricted to a narrower region encompassing the posterior support identified by a Fisher analysis. \ 

Finally, we generate a noiseless\footnote{We follow this approach for visualization purposes, so that the resulting posteriors are centered on the fiducial values.} mock observation using the fiducial parameters in Tab. \ref{tab:priors}. As our baseline configuration, we adopt an $\ell_{\rm max}=3000$ scale cut for the full \threetwopt~data vector. We construct an analytical Gaussian covariance matrix using \texttt{Spaceborne}\footnote{\href{https://github.com/davidesciotti/Spaceborne}{https://github.com/davidesciotti/Spaceborne}} at exactly the same fiducial. In constructing the covariance, we assume a survey area of 14,000~deg$^2$, corresponding to the expected coverage of a Stage-IV photometric survey by the end of the nominal mission \citep{Euclid:2024yrr}.

\begin{table}[h!]
\caption{Priors adopted for this work, together with the fiducial values used to generate the noiseless data vector and corresponding covariance matrix. $\mathcal{U}(a,b)$ denotes a uniform distribution on $[a,b]$ and $\mathcal{N}(\mu,\sigma)$ a Gaussian of mean $\mu$ and standard deviation $\sigma$. Parameters indexed by $i$ are repeated independently for each of the $N_{\rm bin}=13$ tomographic bins. In this work, we fix the baryonic feedback to $\log_{10}T_{\rm AGN}=7.75$.}
\label{tab:priors}
\centering
\begin{tabular}{lcc}
\toprule
Parameter & Prior & Fiducial \\
\midrule
\multicolumn{3}{l}{\textit{Cosmology}} \\
$H_0$ & $\mathcal{U}(50, 90)$ & $67.0$ \\
$\Omega_{b}$ & $\mathcal{U}(0.01, 0.1)$ & $0.049$ \\
$\Omega_{\rm cdm}$ & $\mathcal{U}(0.1, 0.8)$ & $0.27$ \\
$n_s$ & $\mathcal{U}(0.6, 1.2)$ & $0.96$ \\
$\ln(10^{10}A_s)$ & $\mathcal{U}(1.79, 3.91)$ & $3.045$ \\
\midrule
\multicolumn{3}{l}{\textit{Nuisance -- intrinsic alignment}} \\
$A_{\rm IA}$ & $\mathcal{U}(-1, 1)$ & $0.16$ \\
$\eta_{\rm IA}$ & $\mathcal{U}(-5, 5)$ & $1.66$ \\
\midrule
\multicolumn{3}{l}{\textit{Nuisance -- galaxy bias}} \\
$b_{g1}$ & $\mathcal{U}(-3, 3)$ & $1.333$ \\
$b_{g2}$ & $\mathcal{U}(-3, 3)$ & $-0.724$ \\
$b_{g3}$ & $\mathcal{U}(-3, 3)$ & $1.018$ \\
$b_{g4}$ & $\mathcal{U}(-3, 3)$ & $-0.149$ \\
\midrule
\multicolumn{3}{l}{\textit{Nuisance -- per tomographic bin ($i=1,\ldots,13$)}} \\
$b_{{\rm mag},i}$ & $\mathcal{U}(-2, 2)$ & $0$ \\
$m_i$ & $\mathcal{N}(0, 0.01)$ & $0$ \\
$\Delta z_i$ & $\mathcal{N}(0, [0.002(1+\bar{z}_i)])$ & $0$ \\
\bottomrule
\\
\end{tabular}
\end{table}

\section{Simulation-based inference}
\label{sec:sbi}

The main goal of Bayesian inference is the probability distribution of the parameters $\thet$ of a certain cosmological model $\mathcal{M}$ given the observed data $\xx$. According to Bayes' theorem, this probability can be obtained as
\begin{equation}
p(\thet|\xx) = \frac{p(\xx|\thet)\,p(\thet)}{p(\xx)},
\label{eq:bayes}
\end{equation}
where $p(\xx|\thet)$ is called the \emph{likelihood}, $p(\thet)$ is the \emph{prior}, $p(\xx)$ is the \emph{evidence}, and $p(\thet|\xx)$ is known as \emph{posterior}. Classical sampling algorithms, such as MCMC or nested sampling, rely on repeated evaluations of the likelihood to get samples of the full joint distribution $p(\thet|\xx)$. As discussed previously, this becomes computationally expensive as the number of parameters and complexity of the model increases.

In SBI, the information about the likelihood is implicitly accessed via a simulator. By drawing parameters from the prior $p(\thet)$ and calling the simulator,  we can generate samples of data-parameter pairs drawn from the joint distribution $p(\xx,\thet) = p(\xx|\thet) p(\thet)$, which constitute the training data for the inference networks. There are various flavors of SBI, depending on which quantity the network is trained to approximate: Neural Posterior Estimation (NPE), Neural Likelihood Estimation (NLE), and Neural Ratio Estimation (NRE). While both NPE and NLE require estimating a normalized probability density \citep{Alsing:2018eau, Alsing:2019xrx}, NRE reframes inference as a simple binary classification task, allowing much greater flexibility in network architecture. In this work we adopt a variant of NRE known as Marginal Neural Ratio Estimation (MNRE), implemented in the public code \Swyft\footnote{\href{https://github.com/undark-lab/swyft}{https://github.com/undark-lab/swyft}} \citep{Miller:2021hys}.

\subsection{Marginal Neural Ratio Estimation}

Given $N$ data-parameter pairs $\{(\xx^1,\thet^1),...,(\xx^N,\thet^N)\}$ sampled from the joint distribution $p(\xx,\thet)$, a second ensemble drawn from the product of marginals $p(\xx)p(\thet)$ can be constructed simply by randomly shuffling the parameter and data components of the original set. NRE exploits both ensembles to train a network that approximates the ratio
\begin{equation}
r(\xx;\thet) \equiv \frac{p(\xx,\thet)}{p(\xx)\,p(\thet)} = \frac{p(\xx|\thet)}{p(\xx)} = \frac{p(\thet|\xx)}{p(\thet)},
\label{eq:mnre_ratio}
\end{equation}
where the last two equalities follow trivially from  Bayes' theorem in Eq.~\ref{eq:bayes}. Estimating $r(\xx;\thet)$ is therefore equivalent to estimating either the likelihood-to-evidence ratio or the posterior-to-prior ratio, and posterior samples can subsequently be obtained by drawing from the prior $p(\thet)$ and reweighting by $r(\xx;\thet)$.\ 

In practice, $r(\xx;\thet)$ is recovered by training a binary classifier $d_\phi(\xx,\thet)$, with learnable parameters $\phi$, to distinguish jointly drawn pairs from marginally drawn ones,
\begin{equation}
d_\phi(\xx,\thet) \simeq
\begin{cases}
1 & \text{if } (\xx,\thet) \sim p(\xx,\thet), \\
0 & \text{if } (\xx,\thet) \sim p(\xx)p(\thet),
\end{cases}
\end{equation}
by minimizing the binary cross-entropy loss (see \citealt{Miller:2021hys} for details).\ 

A key strength of this approach is that marginal posteriors can be obtained directly, simply by omitting parameters from the network's input -- this is the MNRE variant\footnote{In MNRE the marginalization is implicit, as it results from varying all parameters $\thet$ when generating the simulations.}. Concretely, if $\varthet$ denotes the (typically low-dimensional) subset of parameters of interest, with $\thet = (\varthet,\et)$, the network can be trained to estimate $r(\xx;\varthet) = p(\varthet|\xx)/p(\varthet)$ directly. Here we restrict ourselves to 1- and 2-dimensional marginals, i.e. $\varthet = \theta_k$ for some $k$, or $\varthet = (\theta_i,\theta_j)$ for some $(i,j)$. This is especially useful in cosmology, where scientific interest is usually confined to a small subset of the full parameter space.\ 

Another important aspect of SBI is data compression. When working with high-dimensional data $\xx$ (as in this case, where the data vector has a size $N_{\rm spectra} N_{\rm bin, \ell} = 11232$), it is often advantageous to compress the data before the inference step. Within MNRE, this amounts to estimating $r(\bs,\thet)$, where the lower-dimensional feature vector $\bs = C_\phi(\xx)$ is produced by a compression network $C_\phi$, trained jointly with the ratio-estimation network by minimizing the same loss function. The resulting summaries $\bs$ can be shown to maximize the Jensen--Shannon divergence between the prior and the posterior \citep{Cole:2021gwr}.\ 

In the following subsections, we describe the main steps in our SBI pipeline, which are schematically summarized in Fig.~\ref{fig:pipeline}. 

\begin{figure*}
\centering
\includegraphics[width=0.45\linewidth]{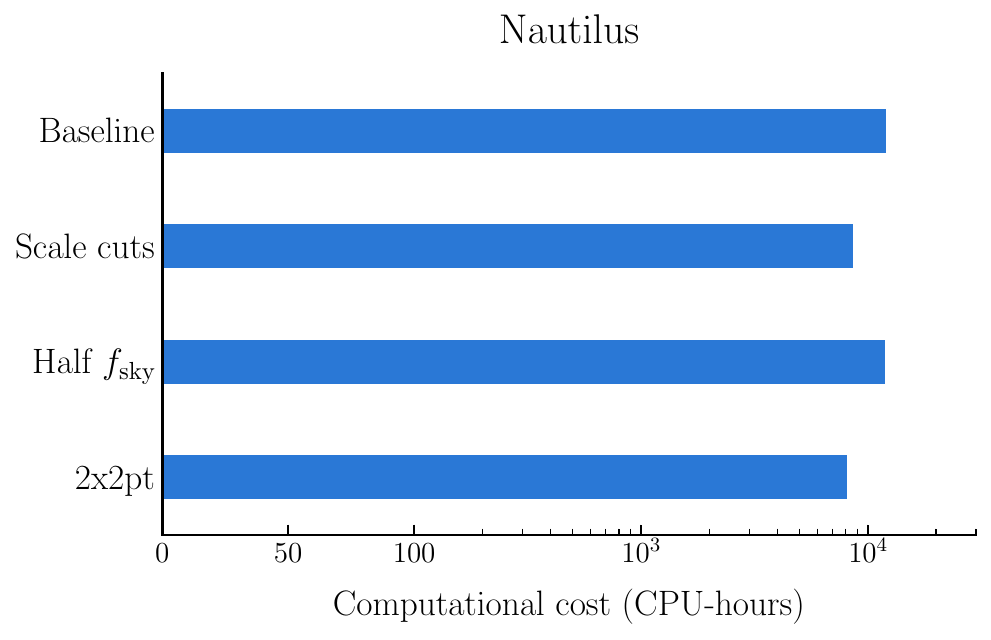}
\hspace{10mm}
\includegraphics[width=0.45\linewidth]{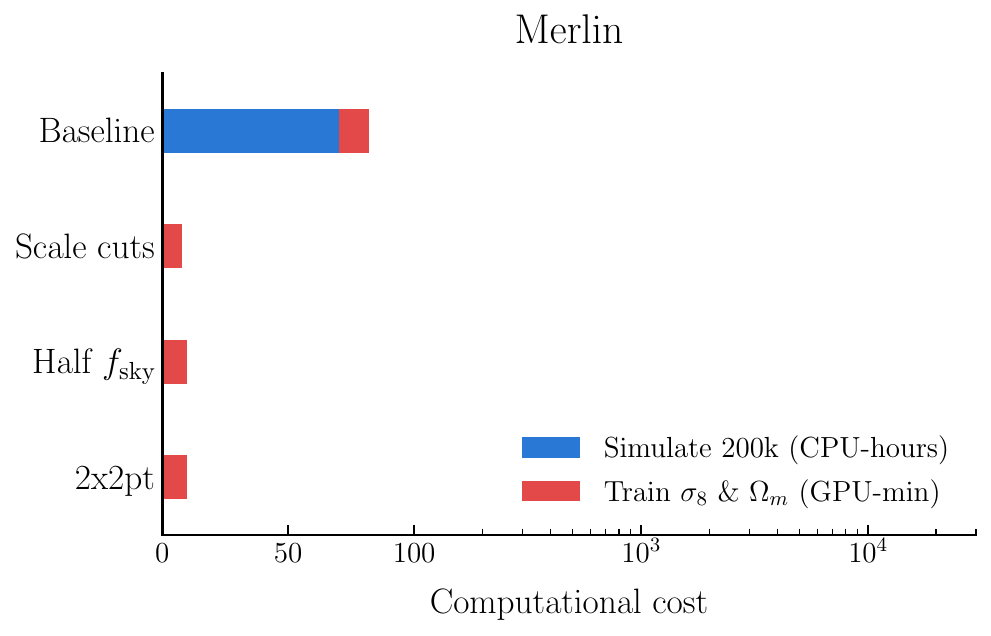}
\caption{Comparison of computational cost between  \Nautilus~and \Merlin~across the analysis variants considered in this work. For \Nautilus, we show the number of CPU-hours needed to reach convergence (typically $\sim4$--$5$~days on 70-100 CPUs, resulting in $\sim \mathcal{O}(10^7)$ likelihood evaluations per run). For \Merlin~we show the CPU-hours needed to generate the 200k training simulations ($\sim1$~hour on 70 CPUs) and the GPU-minutes needed to train the $\sigma_8$ and $\Omega_m$ inference networks ($\sim10$--$15$~min on a single GPU). Because the simulations generated for the baseline \threetwopt~analysis can be reused, the remaining variants only require retraining the networks, at \emph{zero} extra simulator cost. The additional steps required by \Merlin~(e.g., initial Fisher, pre-processing of data, Cholesky rotation, PCA compression) take negligible time by comparison and hence are not shown here. Training for a larger number of parameters, e.g. 10--15, results only in a modest increase in training time (see App.~\ref{app:nuisance}).}
\label{fig:timing}
\end{figure*}

\subsection{Simulation}
\label{sec:simulation}

\subsubsection{Simulator and priors}
\label{sec:fisher}

The simulator-efficiency of SBI methods can be drastically improved when simulations are restricted to a region of parameter space that  resembles the target observation, rather than sampled blindly from the full prior. To address this issue, sequential SBI methods have been developed. For instance, very wide priors can be accommodated in \Swyft~via an extension of the algorithm called Truncated Marginal Neural Ratio Estimation (TMNRE, \citealt{Cole:2021gwr}), which applies a truncation scheme to iteratively zoom into the relevant region of parameter space. Here we opt for a simpler alternative, following \citet{FrancoAbellan:2024tbj}, and estimate the region where
the posterior has support directly from a Fisher forecast.

The Fisher matrix quantifies the local curvature of the log-likelihood around the fiducial parameters $\thet^0$,
\begin{equation}
    F_{\alpha\beta} = -\left. \frac{\partial^2 \ln\mathcal{L}}{\partial\theta_\alpha\partial\theta_\beta}\right\rvert_{\thet^0}\,,
    \label{eq:fisher-def}
\end{equation}
and its inverse provides, in the Gaussian approximation, the formal uncertainty on each parameter $\sigma_i^F = \sqrt{(F^{-1})_{ii}}$. Using the \threetwopt~observables and notation of
Sect.~\ref{sec:cells}, the Fisher matrix can be written as \citep{Euclid:2019clj}
\begin{equation}
\begin{aligned}
F_{\alpha\beta} ={}&
\sum_{\ell=2}^{\ell_{\max}} \sum_{AB,A'B'} \sum_{ijmn}
\frac{\partial C_{ij}^{AB}(\ell)}{\partial\theta_\alpha}
\mathbf{C^{-1}}[C_{ij}^{AB}(\ell), C_{mn}^{A'B'}(\ell)] \\
&\times
\frac{\partial C_{mn}^{A'B'}(\ell)}{\partial\theta_\beta},
\end{aligned}
\label{eq:fisher-full}
\end{equation}
where $\mathbf{C}$ is the data covariance matrix.\ 

We generate $N_{\rm sim} = 2\times 10^5$ simulations in parallel by varying all 5 cosmological and 45 nuisance parameters, where each parameter is drawn from the following prior\footnote{For parameters assigned Gaussian priors in Tab.~\ref{tab:priors}, we instead draw from the corresponding truncated Gaussian, rejecting any sample falling outside the Fisher-restricted region.},
\begin{equation}
    \theta_i \sim \mathcal{U}\!\left(\left[\theta_i^0 - 7\sigma_i^F,\, \theta_i^0 + 7\sigma_i^F\right]\right).
    \label{eq:fisher-prior}
\end{equation}
We have chosen $N_{\rm sim} = 2\times 10^5$ together with a $7\sigma$ margin so that a single simulation bank can resolve the posteriors for all the analysis variants considered in this work. However, we verified that targeting
just the baseline analysis requires only $N_{\rm sim} = 5\times 10^4$ simulations and a $5\sigma$ margin.

Each sampled parameter set is then passed to \cloelib~to compute the corresponding \threetwopt~angular power spectra  (Sect.~\ref{sec:cells}), and a Gaussian noise realisation
$n_{ij}^{AB} \sim \mathcal{N}(\mathbf{0}, \mathbf{C})$ is added to each data vector. 

\subsubsection{Derived parameters}
\label{sec:derived}

In cosmological \threetwopt~analyses it is common to report constraints on derived parameters such as $\sigma_8$ and $\Omega_m$. Inferring these within SBI, however, requires some care: since $\sigma_8$ depends non-linearly on the five \lcdm~ parameters, even the relatively narrow priors of Eq.~\ref{eq:fisher-prior} translate into a derived prior on $\sigma_8$ that is much wider than the small posterior
support expected from a \threetwopt~analysis, leaving few simulations in the region of interest. The situation is easier for $\Omega_m$, since it has a trivial linear dependency on $\Omega_{b}$ and $\Omega_{\rm cdm}$. \

To circumvent this issue, we decided to follow a simple rejection sampling approach. First, we estimate the approximate region of support for  $\theta_i = \{\sigma_8, \Omega_m\}$ as $\left[\theta_i^0 - 7\sigma_i^F,\, \theta_i^0 + 7\sigma_i^F\right]$, where the error $\sigma_i^F$ is obtained by propagating the Fisher covariance of the \lcdm~cosmological parameters through the Jacobian of the transformation. During simulation, cosmologies whose values of $ \sigma_8$ and $\Omega_m$ fall outside this restricted region are rejected.\ 

This procedure can lead to a small acceptance rate and thus to an excessive number of $\sigma_8$ evaluations (which is expensive if one is not using an emulator). To accelerate the rejection sampling, we build a quadratic
Taylor approximation for $\sigma_8$ around the fiducial cosmology, and use it only to reject points that lie clearly outside the restricted region, while still performing the full evaluation for points near or inside it.

\subsection{Data compression and inference networks}

\subsubsection{Initial pre-compression}

The simulated data vectors contain a large number of correlated noisy power spectra, so it is beneficial to first perform a Cholesky decomposition of the covariance matrix $\mathbf{C} = \mathbf{L}\mathbf{L}^{\top}$ (where $\mathbf{L}$ is a lower triangular matrix) and then change to the basis
\begin{equation}
    \tilde{\xx} = \mathbf{L}^{-1}\xx ,
\end{equation}
in order to obtain noise-whitened data vectors. Then, we do a Principal Component Analysis (PCA) in order to rotate these whitened simulations to a smaller basis capturing the largest data variations.  We retain only the first $N_{\rm PCA}$ components whose singular value contribute to more than $\epsilon_{\rm v} = 0.001\%$ of the total (typically $N_{\rm PCA} \sim 300-400$).\footnote{We note that this threshold $\epsilon_{\rm v}$ is smaller than the one used in \citet{FrancoAbellan:2024tbj}, which marginalized over less nuisance parameters and thus had fewer data variations and fewer PCA components to retain.}

This pre-compression step needs to be performed only once at the
beginning, before training, and helps to lower the required capacity of the neural compression network described below.

\subsubsection{Network architecture and training}

The design of the network is split into two parts:
\begin{enumerate}
    \item A \emph{compression head}: a multi-layer perceptron (MLP) mapping the PCA-projected, Cholesky-whitened data vector (passed through an online
    normalisation layer) to a feature vector of size $N_{\rm par} \times N_{\rm feat}$, via hidden layers of dimension $2048 \to 512$. Here $N_{\rm par}$ is the number of parameters of interest, and $N_{\rm feat}$ denotes a fixed number of features per parameter (typically $N_{\rm feat} = 1$--$4$).
    
    \item A set of \emph{ratio estimators}: each parameter of interest $\theta_i$ (for $i=1,..,N_{\rm par}$) is associated with its own feature vector $\bs_i$ (of size $N_{\rm feat}$), which are fed as input to the ratio estimators $r(\bs_i;\theta_i)$ and $r(\bs_i, \bs_j;\theta_i,\theta_j)$ in charge of estimating every 1- and 2-dimensional marginal posterior. The ratio estimators are implemented as MLPs consisting of four residual blocks with 128 neurons each and a dropout $p=0.1$.
    \end{enumerate}

All networks are trained with the \texttt{AdamW} optimiser together with a \texttt{ReduceLROnPlateau} scheduler that reduces the learning rate (lr) by a factor of 10 whenever the validation loss fails to improve for 3 consecutive epochs. Training is stopped early if the validation loss does not improve for 5 epochs, and the checkpoint with the lowest validation loss is kept. We use a batch size of 256, an initial learning rate of $\mathrm{lr} = 5\times 10^{-4}$, and reserve $20\%$ of the simulated samples for validation.\ 

During training, noise realisations are independently resampled for each spectrum at every iteration, since this allows to prevent overfitting. 

\subsection{Inference}
\label{sec:inference}

Once trained, the ratio estimators give direct access to
$r(\xx;\thet)$ for all trained parameters. Marginal 1- and 2-dimensional posteriors are obtained by quickly
evaluating the corresponding estimator on a dense grid of prior
samples and the observation $\mathbf{x}_{0}$, and weighting each sample by $r(\xx_0;\thet)$. Since this inference is locally amortized, the same trained networks can sample from the marginal posteriors for any observation $\xx$ drawn from the (Fisher-restricted) prior region. This allows for useful cross-checks such as coverage tests, which we present in
App.~\ref{app:coverage}.

When performing inference for derived parameters like $\sigma_8$, the prior has no closed analytical form. Hence, for these parameters we evaluate the trained $r(\xx;\thet)$ directly on samples from the simulation store, which follow the correct prior by construction.

\section{Results}
\label{sec:results}

\subsection{Baseline posteriors}
\label{sec:posteriors}

We validate \Merlin~against the nested sampler \Nautilus~\citep{Lange:2023ydq}, using the \texttt{cloelike}\footnote{\href{https://github.com/cloe-org/cloelike}{https://github.com/cloe-org/cloelike}} likelihood module and an otherwise identical $3\times2$pt analysis setup (Sect.~\ref{sec:cells}), ensuring a consistent comparison. In Fig.~\ref{fig:timing}, we benchmark the computational performance of both approaches. \ 

\begin{figure}[h!]
\includegraphics[width=0.4\textwidth]{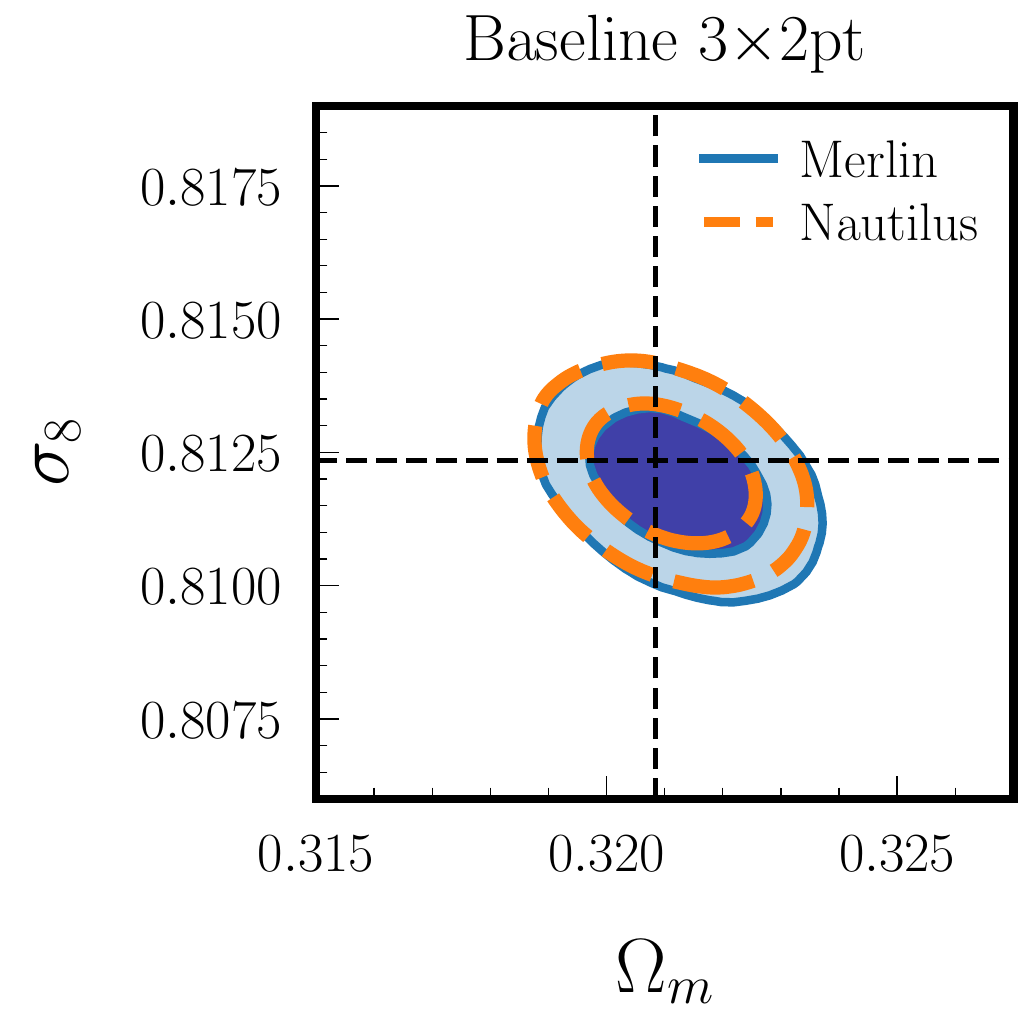}
\caption{2-dimensional marginalized posterior distribution for $\sigma_8-\Omega_m$ obtained with \Merlin~(blue) and \Nautilus~(orange). The dashed lines indicate the fiducial cosmology. \Merlin~is in excellent agreement with \Nautilus~while requiring substantially less computational time.}
\label{fig:baseline}
\end{figure}

We show constraints on $\sigma_8$ and
$\Omega_m$ in Fig.~\ref{fig:baseline}, obtained by marginalising over the 5 \lcdm~cosmological parameters and the 45 nuisance parameters (intrinsic alignments, galaxy bias, magnification bias, multiplicative shear calibration, and photometric-redshift uncertainties). We can see that the \Merlin~and \Nautilus~posterior contours are in excellent agreement.  

The computational cost of the two approaches differs substantially. For \Merlin, generating the $N_{\rm sim} = 2 \times 10^5$ simulations required  $\sim$1 hour on 70 CPU cores, followed by $\sim$10 minutes of network training on a single 94 GB \texttt{Nvidia H100} GPU. By comparison, the \Nautilus~analysis required $\sim$5 days on 100 CPU cores (i.e.~$\sim$$10^4$ CPU-hours). This massive reduction in computational time stems from two factors. First, \Merlin~directly estimates the marginal posteriors $p(\theta_i|\xx)$ and $p(\theta_i, \theta_j |\xx)$, which requires far fewer simulations than first sampling the full 50-dimensional joint distribution $p(\thet|\xx)$ and then marginalising.\footnote{The full set of cosmological and nuisance parameters can nevertheless be obtained by training from the same simulation bank, and their posterior constraints are also consistent with those obtained from nested sampling (see App.~\ref{app:nuisance}).} In particular, achieving converged posteriors for nested sampling required $\mathcal{O}(10^7)$ likelihood evaluations, a factor of $\mathcal{O}(10^2)$ more than $N_{\rm sim}$. Second, the generation of simulations for SBI is embarrassingly parallel, a property we exploit throughout our pipeline. This comparison demonstrates that \Merlin~can reproduce conventional posterior constraints at a substantially reduced computational cost.

\subsection{Analysis variants}
\label{sec:variants}

We now leverage the simulation reusability of SBI to perform three re-analyses without having to generate a new training set. This contrasts with conventional sampling-based analyses, for which each modification to the analysis setup would require an entirely new run. We select three representative use cases that commonly arise in LSS analyses and forecasts, where changes to the analysis setup are independent of the underlying simulator model and hence can be achieved by pre-processing the existing simulation bank. Specifically, we consider:

\begin{figure*}[ht!]
\centering

\includegraphics[width=0.32\linewidth]{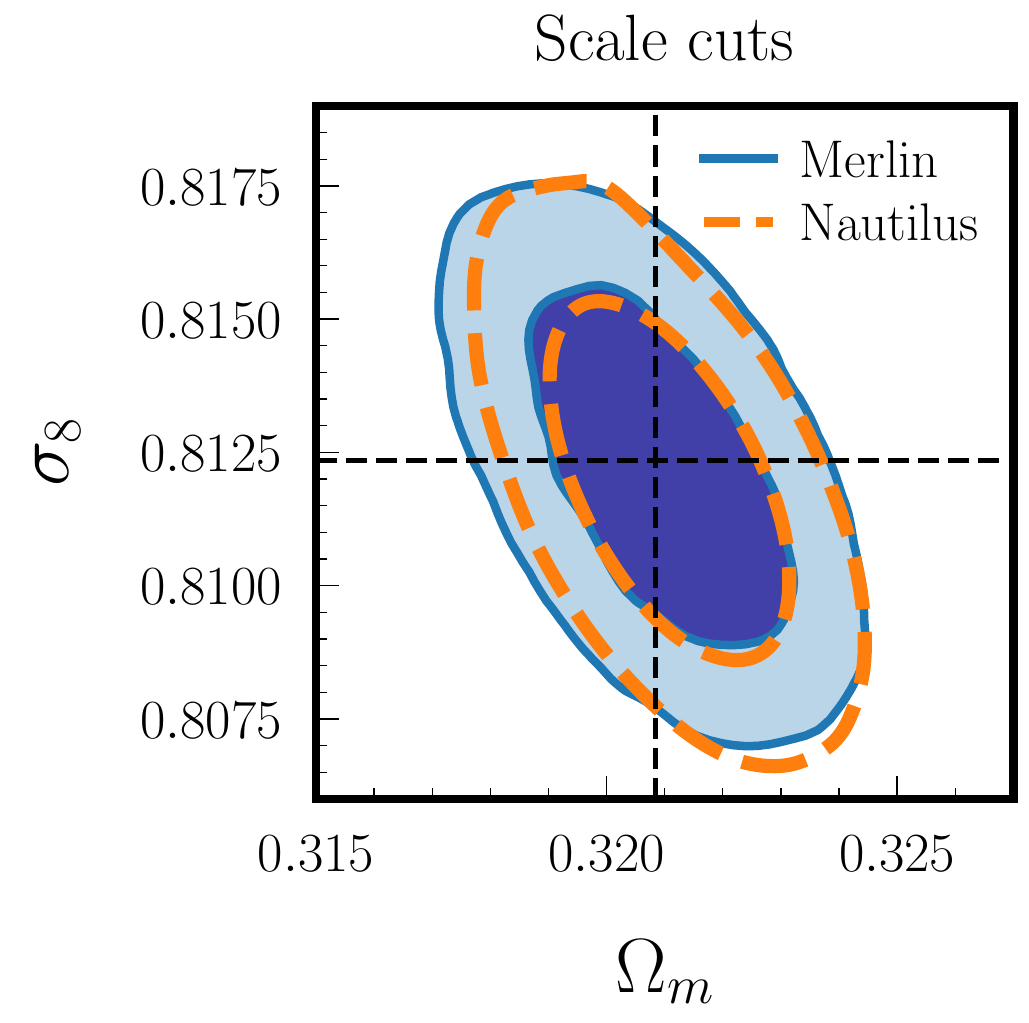 }
\hspace{1mm}
\includegraphics[width=0.32\linewidth]{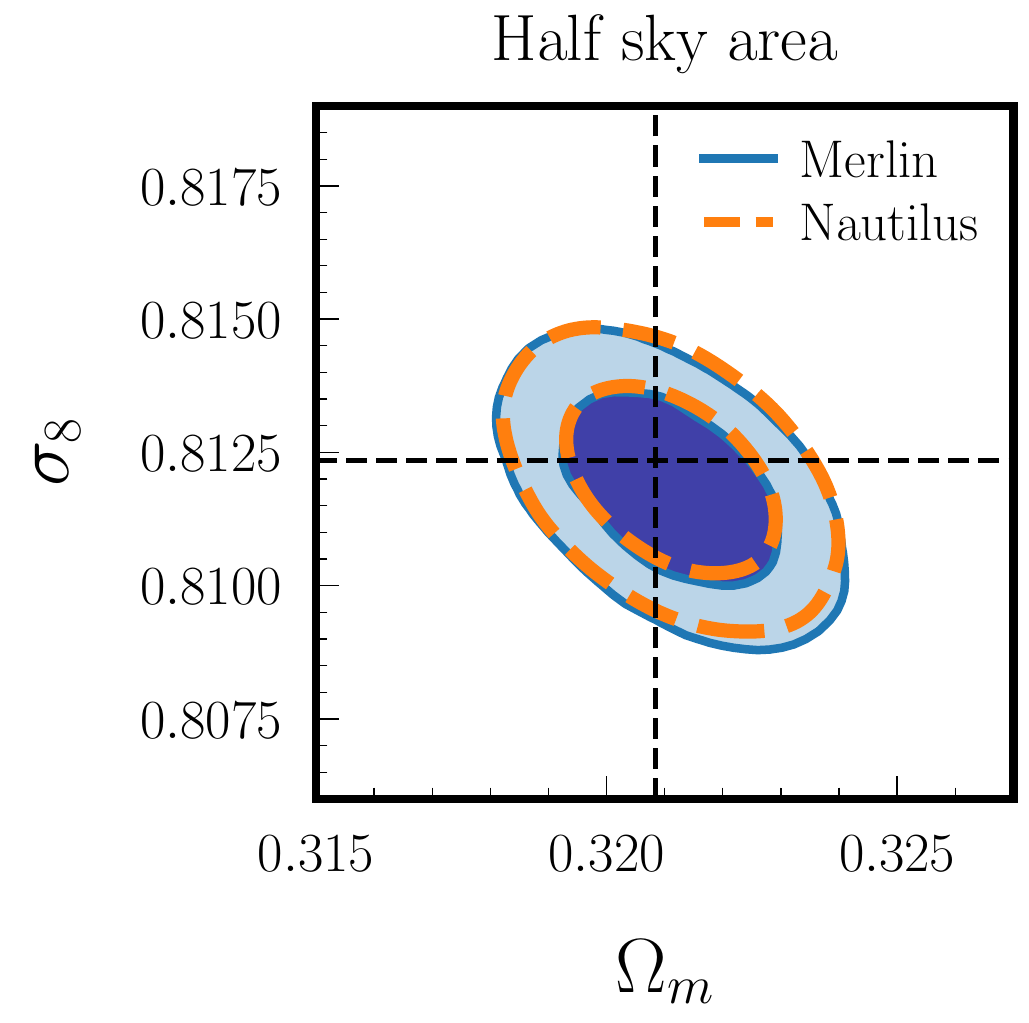}
\hspace{1mm}
\includegraphics[width=0.32\linewidth]{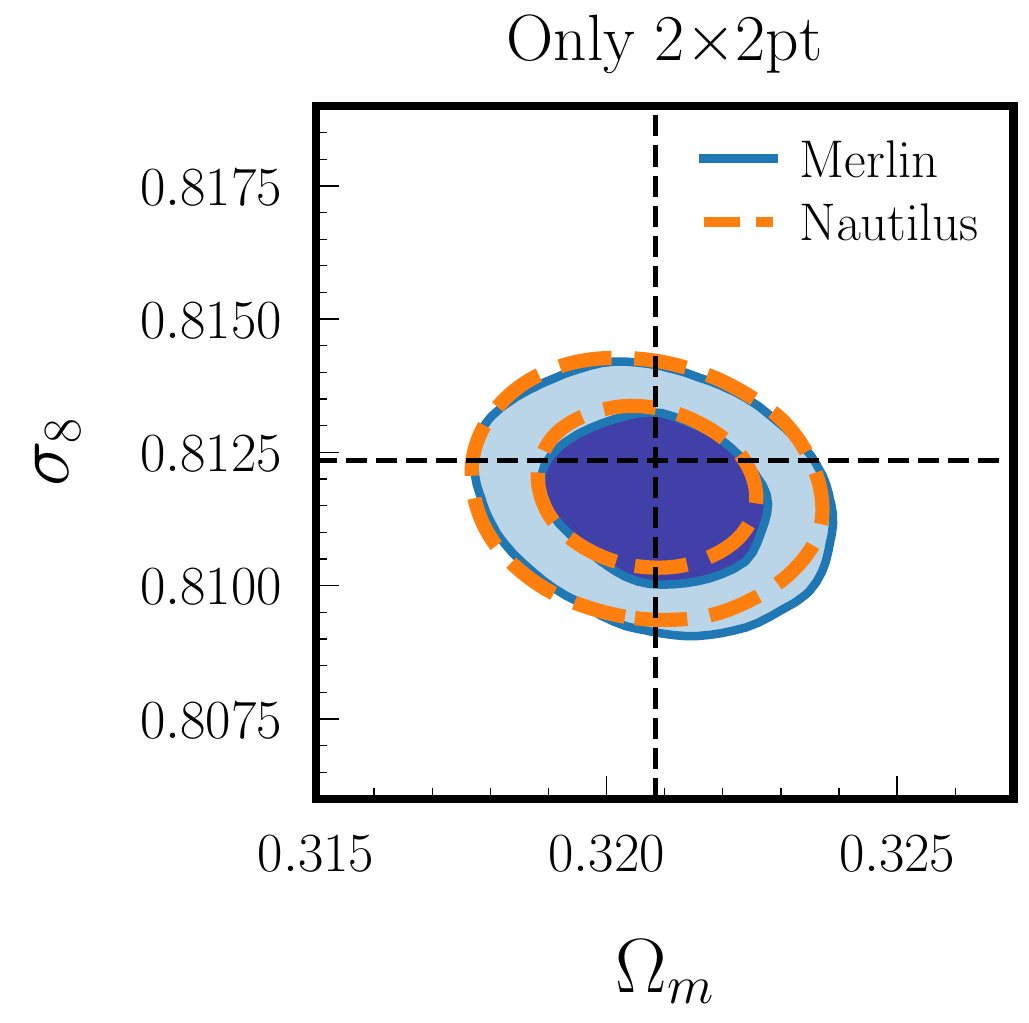 }
\caption{Posterior distributions for $\sigma_8-\Omega_m$ obtained under several variants in the analysis. \emph{Left panel}: scale cuts in the data vector ($\ell_{\rm max}^{\rm WL}=1500$, $\ell_{\rm max}^{\rm GC}=\ell_{\rm max}^{\rm XC}=750$). \emph{Middle panel}: a different covariance matrix generated with half $f_{\rm sky}$ ($7000~\rm{deg}^2$). \emph{Right panel}: the combination of only  angular clustering and galaxy-galaxy lensing (2$\times$2pt). All the \Merlin~analyses (blue) reuse the same simulation store, only requiring to retrain the networks, while \Nautilus~(orange) requires separate runs for each case. 
}
\label{fig:analysis_variants}
\end{figure*}

\begin{itemize}
\item \textbf{Scale cuts}. Scale cuts are an essential component of LSS analyses, as they allow us to determine the range of scales over which the adopted theoretical model provides a sufficiently accurate description of the data, while balancing the extra constraining power provided by smaller scales against potential modelling uncertainties. In Sect.~\ref{sec:scalecuts}, we show that masking high-$\ell$ bins in the existing simulation store and retraining  the networks reproduces the expected broadening of the posterior relative to the full-$\ell_{\rm max}$ analysis. 

\item \textbf{Survey area}. Varying the survey area is useful for understanding the impact of survey specifications and footprint or sky-coverage decisions on the resulting constraints. In Sect.~\ref{sec:covmat}, we show that by drawing new noise realisations from a  covariance matrix corresponding to a reduced sky fraction $f_{\rm sky}$, we can propagate the expected increase in posterior width without having to recompute the \threetwopt~predictions.

\item \textbf{Data combinations}. Comparing different probe combinations is particularly valuable for assessing the relative impact of different observables and identifying which systematic uncertainties can become dominant in a given analysis.  In Sect.~\ref{sec:subset}, we show that restricting the data vector to a subset of probes -- e.g.\ moving from the full \threetwopt~analysis to a 2$\times$2pt combination -- isolates their individual constraining power without having to generate new simulations.

\end{itemize}

These three use cases therefore cover complementary aspects of LSS analysis design: \emph{scale cuts} determine the range over which the theoretical modelling can be trusted, \emph{survey area} quantifies the impact of survey specifications and footprint or coverage choices, and \emph{probe combinations} help assess the relative importance of different observables and their associated systematic uncertainties. The summary of all the cases at the level of $\sigma_8$ and $\Omega_m$ can be found in Fig. \ref{fig:analysis_variants}. In all cases, the expensive simulation step is performed only once, and the resulting simulations are reused across analysis variants. Inference with \Merlin~additionally requires training, which typically takes just a few GPU-minutes. This illustrates the flexibility our proposed SBI approach.

\subsubsection{Scale cuts}
\label{sec:scalecuts}

Scale cuts are routinely adopted in WL analyses to mitigate the impact of small-scale astrophysical and modelling uncertainties, including baryonic feedback and non-linear structure formation. We adopt conservative cuts of $\ell_{\rm max}=1500$ for the WL spectra and $\ell_{\rm max}=750$ for the GGL and GCph spectra.

The scale cuts are implemented by applying a binary mask to both the training set and to the mock observation, setting to zero the $\ell$-bins above the corresponding cutoff before compression. Importantly, no new forward simulations are required: the masked training set is constructed directly from the existing simulation bank.

The resulting posterior distributions are shown in the left panel of Fig.~\ref{fig:analysis_variants}. As expected, removing the high-$\ell$ data points leads to significantly broader posterior distributions than those shown in Fig.~\ref{fig:baseline}. The agreement between the two approaches illustrates a key advantage of \Merlin: a single simulation store can be reused to explore many different scale-cut choices, with the small additional computational cost limited to retraining the inference networks.

\subsubsection{Survey area}
\label{sec:covmat}

We next demonstrate how a change in the effective survey area can be implemented simply by changing the simulator's noise model. To do so, we replace the fiducial Gaussian covariance matrix with one having half the sky fraction $f_{\rm sky}$, corresponding to 7,000 deg$^2$, which increases the noise level and therefore reduces the expected constraining power.

Rather than regenerating the expensive theory predictions, we reuse the existing simulation bank of \threetwopt~spectra and simply draw new $N_{\rm sim}$ noise realisations from the updated covariance matrix, which is computationally cheap since the noise is Gaussian. The networks are then retrained on this new set of noisy simulations. 

The central panel of Fig.~\ref{fig:analysis_variants} shows the resulting contours for the two approaches. The posteriors broaden roughly by a factor $\sqrt{2} \simeq 1.4$ relative to the baseline, as expected from halving the sky area. This demonstrates how \Merlin~allows a change in the survey area — and thus in the noise model — to be propagated without requiring any new signal simulations.

\subsubsection{Data combinations}
\label{sec:subset}
The final demonstration considers a change in the set of probes included in the analysis. Starting from the full \threetwopt~combination, we construct a 2$\times$2pt analysis by restricting the data vector to a subset of the original data vector. This provides a direct test of whether the existing simulation store can be reused to assess the constraining power of different data combinations.\footnote{We note that simulation reuse is not always possible in this
context: a WL-only analysis, for instance, has significantly less constraining power, leading to posteriors much wider than the Fisher-restricted priors of Eq.~\ref{eq:fisher-prior}. In such cases, we recommend resimulating with adapted Fisher priors, which in any case remains substantially faster than a full \Nautilus~run.}

Similar to  the scale-cut analysis, we only have to restrict the original training set and the mock observation, and retrain the inference networks with the 2$\times$2pt data vector. The resulting posterior constraints are shown in the right panel of Fig.~\ref{fig:analysis_variants}, showing that \Merlin~can isolate the contribution of the reduced probe combination without requiring a new simulation campaign.

\section{Conclusions}
\label{sec:concl}

In this work, we have developed an updated SBI  approach to efficiently perform \threetwopt~cosmological analysis. We have introduced \Merlin, a modular pipeline that combines Marginal Neural Ratio Estimation in \Swyft~with the \cloelib~theory model. We demonstrate the pipeline on a realistic setup expected from a  Stage-IV photometric survey by the end of its operations, with a 45-dimensional nuisance parameter space that includes intrinsic alignments, galaxy bias, magnification bias, multiplicative  shear-calibration, and  photometric-redshift uncertainties.

First, we have shown that \Merlin~delivers posterior constraints that are in excellent agreement with the \Nautilus~nested sampler at substantially reduced computational cost. For the setup considered here, the generation of simulations required two orders of magnitude fewer CPU-hours than \Nautilus, plus a few GPU-minutes for training the networks. This computational advantage arises from two factors: i) MNRE directly targets the marginal posteriors of interest, without having to sample the full joint posterior, which makes it scalable to high-dimensional problems; and ii) simulations for SBI can be fully generated in parallel.

The main advantage of our approach, however, extends beyond the speed of a single inference. Once the simulation store has been generated, it can be reused to investigate different analysis choices without repeating expensive model evaluations. We demonstrated this by modifying the scale cuts, changing the survey area, and restricting the data vector to 2$\times$2pt probes, all at \emph{zero} extra simulator cost. This flexibility is particularly valuable for Stage-IV LSS analyses, where choices such as scale cuts, probe combinations, and survey configuration may need to be explored repeatedly for many different cosmological models.

Our results also illustrate an important distinction between SBI and likelihood-based methods. In a likelihood-based analysis, simulation and inference are intertwined processes, so any change to the analysis generally requires a new full exploration of the parameter space. In SBI, by contrast, simulation and inference are distinct steps, so once a simulation bank has been generated, it can be reused for different purposes without repeated model evaluations. This makes SBI particularly attractive not only as an alternative inference method, but also as a framework for systematically exploring different survey configurations and analysis choices and their impact on cosmological constraints.

We envision several developments before \Merlin~can be applied to real data from a full Stage-IV multi-probe experiment. On the data side, we plan to extend the framework to jointly analyse photometric and spectroscopic surveys, and to benchmark its performance against scalable MCMC methods such as HMC. On the modelling side, we intend to consider more complex simulators, capable of forward-modelling systematic effects that cannot easily be captured in an explicit likelihood. Looking further ahead, we plan to apply \Merlin~to multiple beyond-\lcdm~cosmologies, testing its robustness on models that introduce complex parameter degeneracies and highly non-Gaussian posteriors.

Overall, our results support that MNRE provides a practical and computationally efficient framework for cosmological inference at the level of summary statistics. The combination of accuracy, scalability, and simulation reuse makes \Merlin~a promising tool for the increasingly complex landscape of Stage-IV galaxy surveys.

\section*{Author Contributions}
\begin{itemize}
    \item \textbf{Alexandra Wernersson}: Conceptualization, Methodology, Software (led the initial design and implementation of \Merlin), Investigation (explored the network architecture, produced the first results and figures).  Writing -- Review \& Editing.
    \item \textbf{Guillermo Franco-Abell\'an}: Conceptualization, Methodology, Software (redesigned and extended \Merlin, implemented rejection sampling), Investigation, Validation (explored the network architecture, ran the final results presented in this work), Visualization (produced the figures), Supervision, Writing -- Review \& Editing.
    \item \textbf{Guadalupe Ca\~nas-Herrera}: Conceptualization, Methodology (defined the LSS analysis cases), Resources (provided expertise and access to \cloelib~and \texttt{cloelike}), Software (contributed to the design of \Merlin), Investigation (provided validation \texttt{Nautilus} runs), Supervision, Writing -- Original Draft, Writing -- Review \& Editing.
\end{itemize}

\section*{Acknowledgments}
We acknowledge the use of software developed and maintained by the \texttt{cloe-org} team, an open-source ecosystem for cosmological modelling and statistical inference for large-scale structure that has been adopted by the Euclid Consortium. This work used \texttt{cloelib} and \texttt{cloelike}, which are publicly available at
\href{https://github.com/cloe-org}{https://github.com/cloe-org}.
GFA gratefully acknowledges the computer resources at Artemisa, funded
by the European Union ERDF and Comunitat Valenciana as well as the technical support provided by the Instituto de Fisica Corpuscular, IFIC (CSIC-UV). GCH acknowledges that this project is part of the project UNICORN with file number VI.Veni.242.110 of the research programme Talent Programme Veni Science domain 2024 which is (partly) financed by the Dutch Research Council (NWO) under the grant \url{https://doi.org/10.61686/ZCPQI32997}. GCH acknowleges the use of the Leiden Observatory \texttt{ketelmeer} server, funded thanks to the European Research Council (ERC) under the
European Union’s Horizon 2020 research and innovation program (Grant Agreement No. 101053992.

\textbf{Use of Generative AI.} Anthropic's Claude was used to assist with English grammar and stylistic editing of the manuscript, as well as to support the design of the \Merlin~code. All scientific and methodological decisions, implementation choices, analyses, validation, interpretation, and conclusions were developed and verified by the authors, who retain full responsibility for the contents of this work. 

\vspace{7mm}

\bibliographystyle{apsrev4-1}
\bibliography{oja_template}

\begin{appendix}

\section{Full marginalized posterior distributions}\label{app:nuisance}

The results presented in Sect.~\ref{sec:results} focus on the marginal posteriors of $\sigma_8$ and $\Omega_m$, which are  the two cosmological parameters of primary interest for LSS studies. Although MNRE gives us the flexibility to ignore large numbers of nuisance parameters, the same underlying simulations can always be reused to obtain their marginal posteriors as well. Here we show that inferring marginals for the full set of 50 cosmological and nuisance parameters requires only a modest increase in training time relative to the $\sigma_8$--$\Omega_m$ networks described in Sect.~\ref{sec:posteriors} (see Fig.~\ref{fig:timing}).

\begin{figure*}[ht!]
\centering
\includegraphics[width=0.99\textwidth]{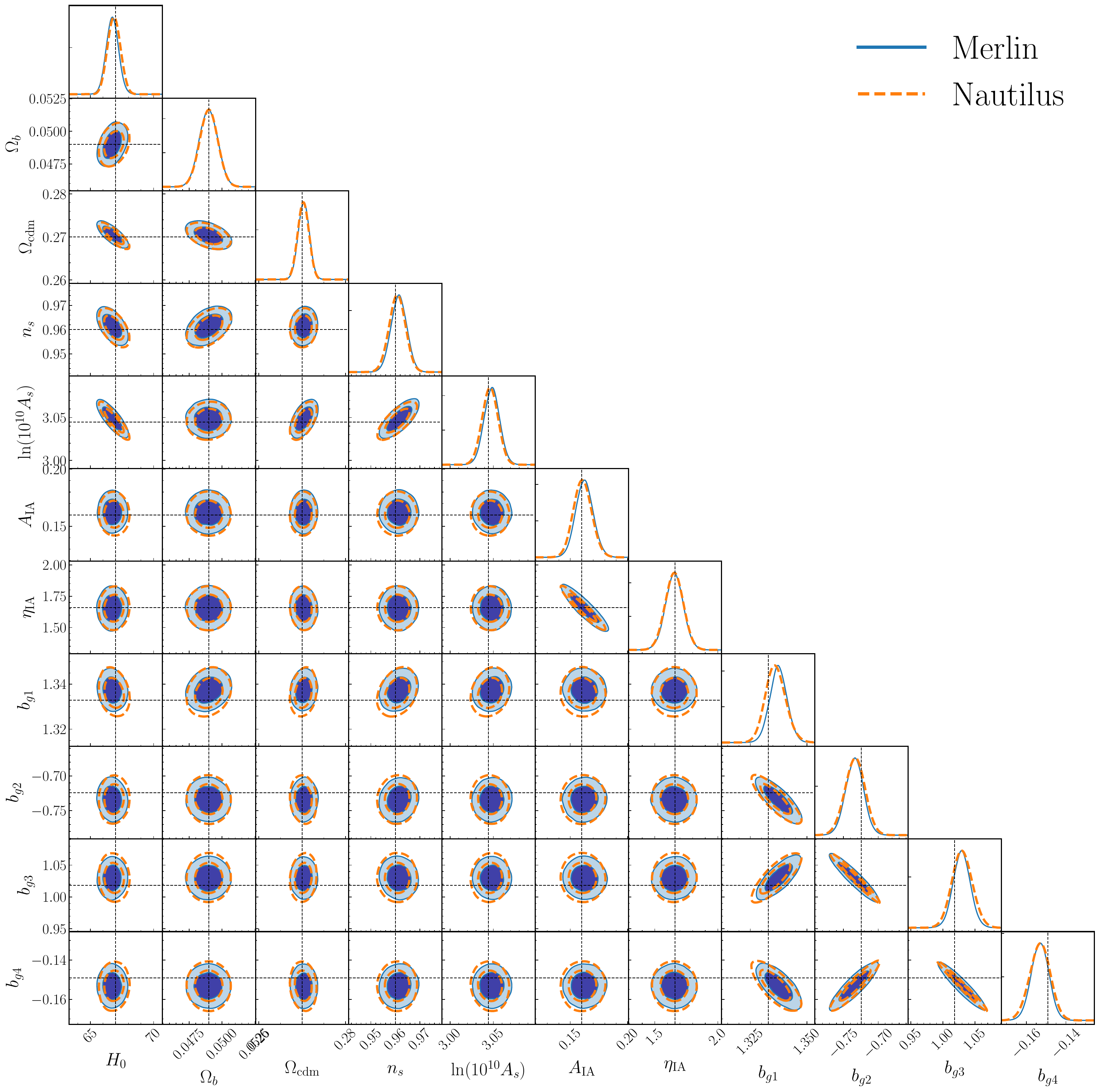}
\caption{Posterior distributions for the five cosmological parameters, the two IA parameters, and the four galaxy bias, obtained with \Merlin~(blue) and \Nautilus~(orange), for the \threetwopt~analysis with reduced sky fraction.}
\label{fig:COSMO_IA_bias}
\end{figure*}

\begin{figure*}[ht!]
\centering
\includegraphics[width=0.99\textwidth]{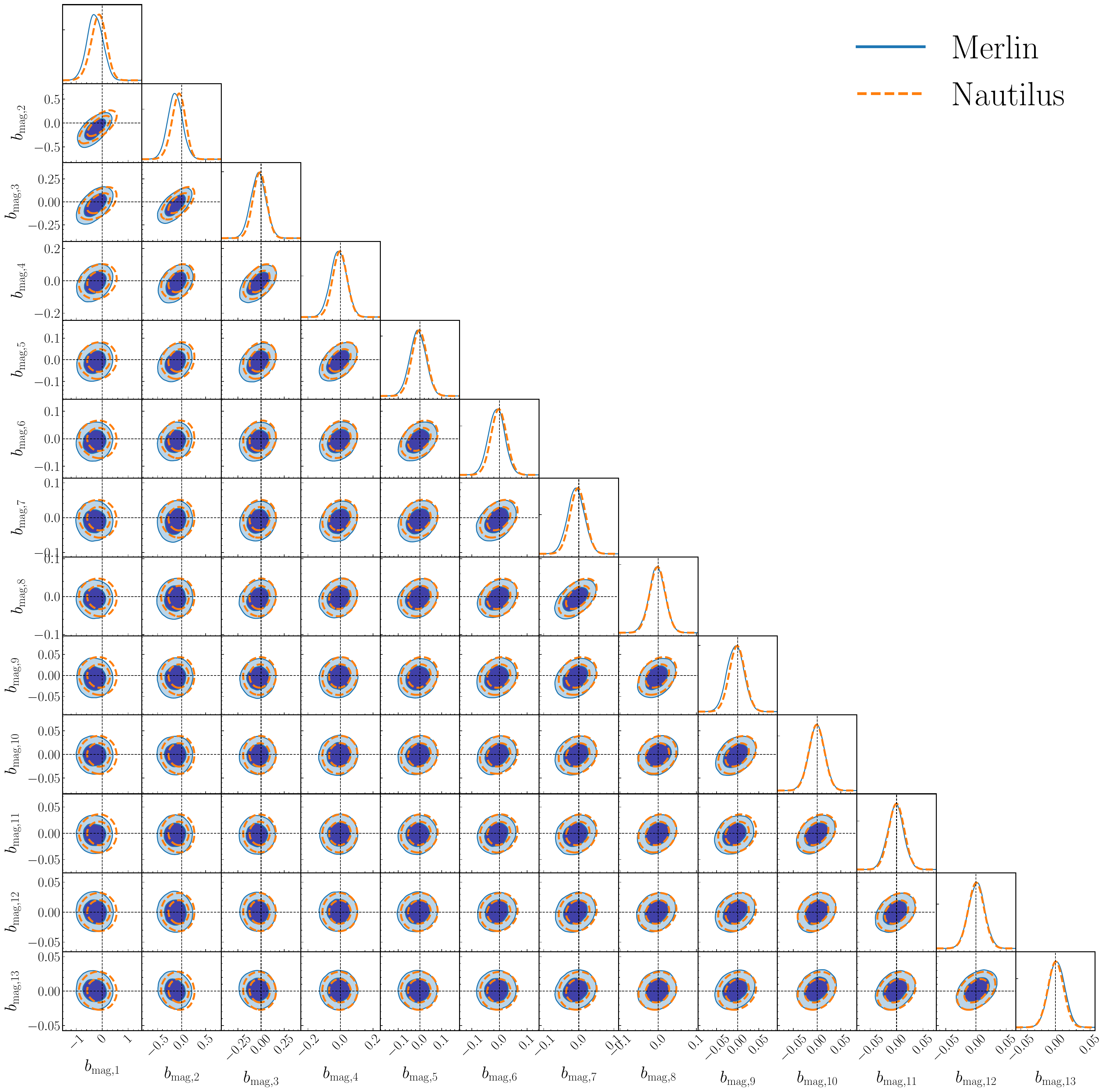}
\caption{Posterior distributions for the 13 magnification bias parameters, obtained with \Merlin~(blue) and \Nautilus~(orange), for the \threetwopt~analysis with reduced sky fraction.}
\label{fig:b_mag}
\end{figure*}

\begin{figure*}[ht!]
\centering
\includegraphics[width=0.99\textwidth]{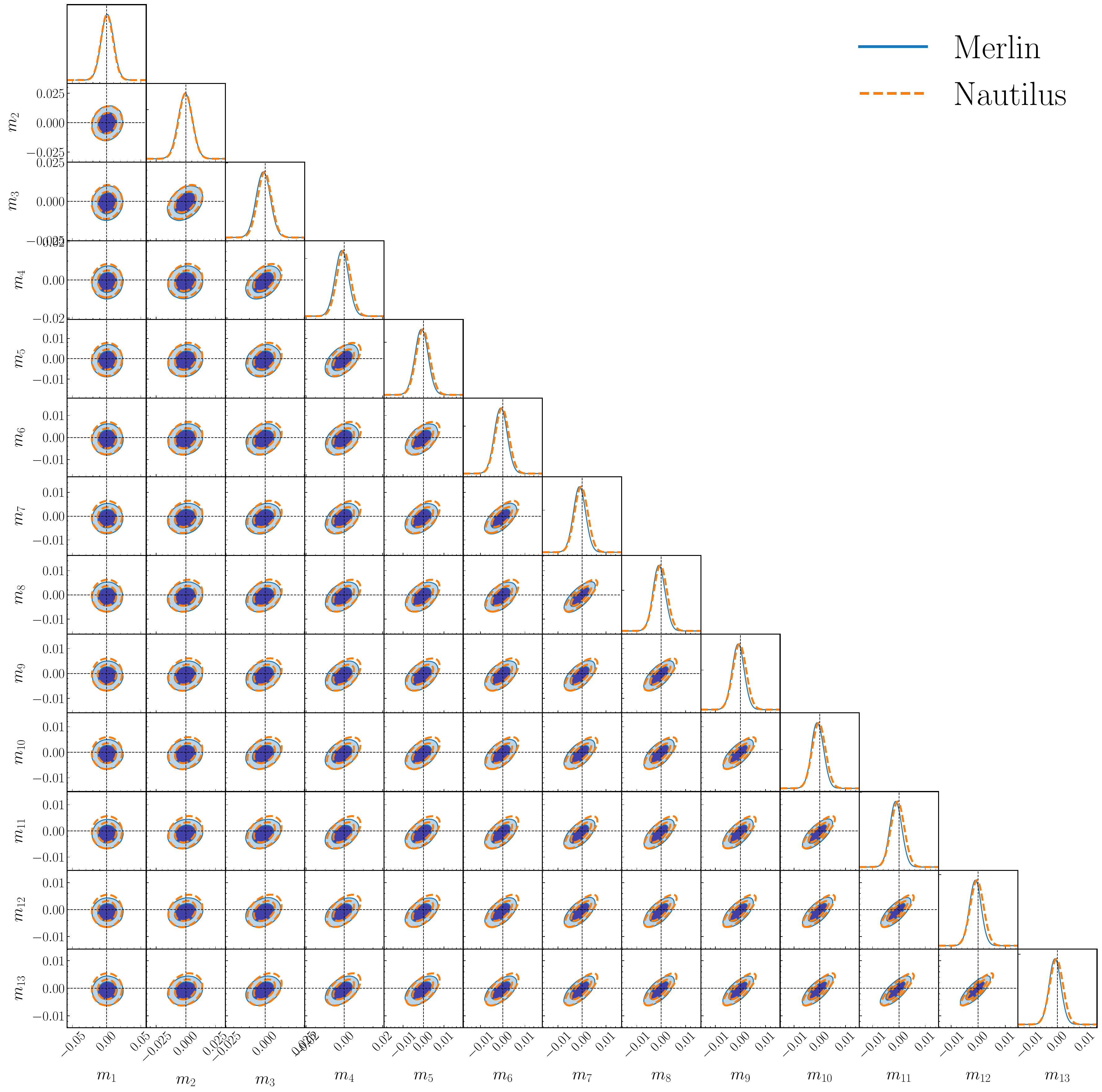}
\caption{Posterior distributions for the 13 multiplicative bias parameters, obtained with \Merlin~(blue) and \Nautilus~(orange), for the \threetwopt~analysis with reduced sky fraction.}
\label{fig:shear_calib}
\end{figure*}

\begin{figure*}[ht!]
\centering
\includegraphics[width=0.99\textwidth]{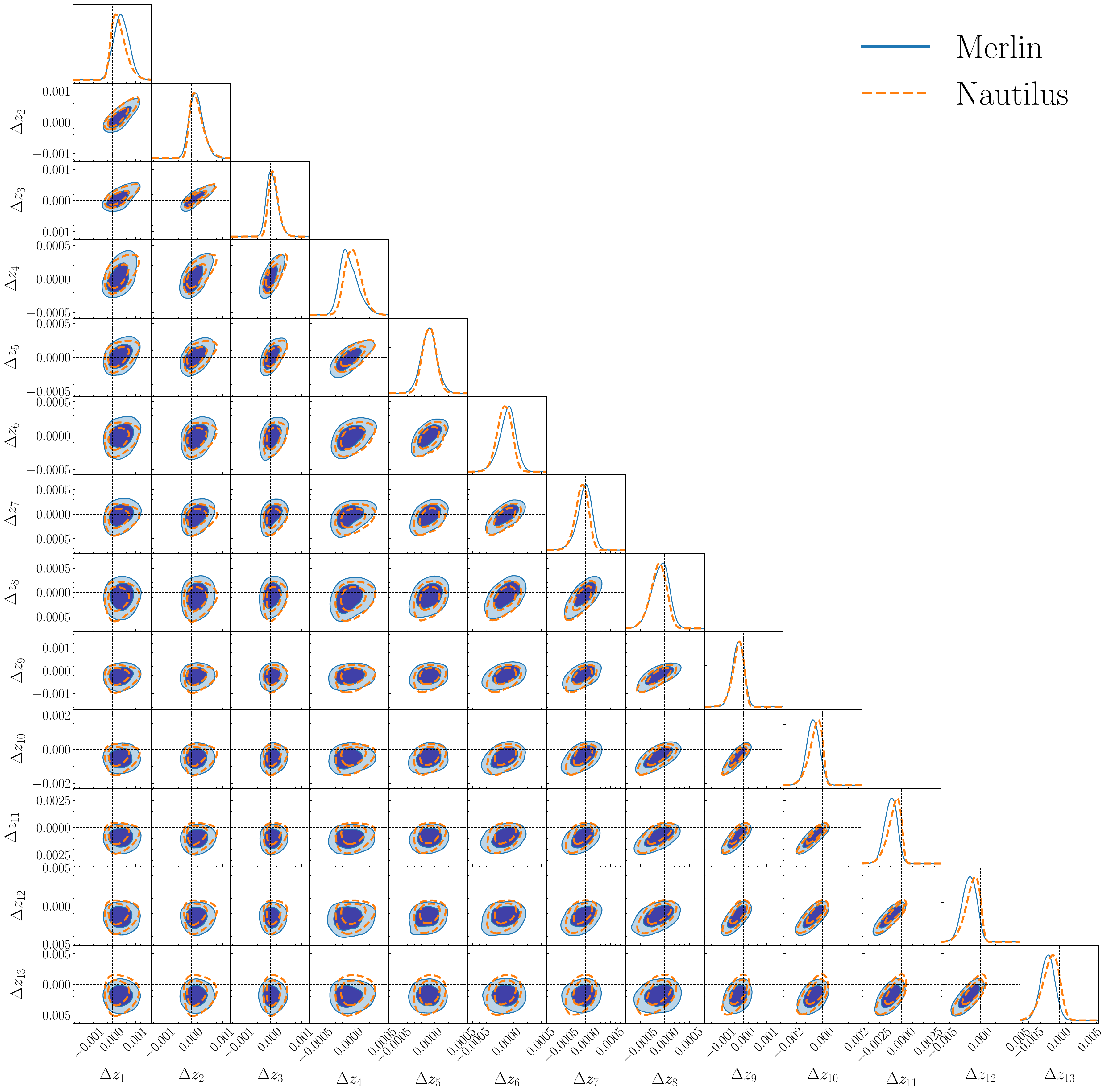}
\caption{Posterior distributions for the 13 photometric-redshift shift parameters, obtained with \Merlin~(blue) and \Nautilus~(orange), for the \threetwopt~analysis with reduced sky fraction.}
\label{fig:photoz}
\end{figure*}

Figs~\ref{fig:COSMO_IA_bias}--\ref{fig:photoz} present the marginal posteriors obtained for the reduced sky-fraction variant described in Sect.~\ref{sec:covmat}, comparing \texttt{Merlin} (blue) against the corresponding \Nautilus~benchmark (orange). Fig.~\ref{fig:COSMO_IA_bias} shows the five cosmological parameters together with the two intrinsic-alignment parameters ($A_{\rm IA}$, $\eta_{\rm IA}$) and the four polynomial galaxy-bias coefficients ($b_{g,0},\ldots,b_{g,3}$); training the corresponding compression--ratio networks required $\sim65$~GPU-minutes. Fig.~\ref{fig:b_mag} shows the 13 tomographic-bin magnification-bias parameters $b_{{\rm mag},i}$, for which training required $\sim100$~GPU-minutes. Fig.~\ref{fig:shear_calib} presents the 13 multiplicative shear-calibration parameters $m_i$, trained in  $\sim 90$~GPU-minutes, while Fig.~\ref{fig:photoz} shows the 13 photometric-redshift shift parameters $\Delta z_i$, with a training time of  $\sim 120$~GPU-minutes. In all cases, the increase in training time relative to the 10--15~GPU-minute reported in Fig.~\ref{fig:timing} is driven by the larger number of  ratio estimators that need to be simultaneously trained. 

Across all four figures, the \texttt{Merlin} posteriors are in good overall agreement with those obtained from \Nautilus, both in terms of their central values and posterior widths, despite strong degeneracies among several of these parameters. Agreement for these parameters therefore provides a particularly stringent test of the inference pipeline.

\section{Coverage tests}
\label{app:coverage}

One of the key motivations for using SBI is to tackle problems where classical inference is prohibitively slow or simply unfeasible, so it is important to have consistency checks that do not rely on a ground-truth reference (such as a converged MCMC run). A key advantage of amortized SBI methods is that, once trained, the inference network can evaluate the posterior not only for the actual observation --- as MCMC does --- but for any mock observation drawn from the prior. This makes it possible to assess the statistical calibration of \Merlin~through coverage tests, without the cost of running a separate chain for every mock observation, as an analogous test would demand for MCMC \citep{Hermans:2021rqv}.

To perform this test, one draws a large number of parameter samples from the prior, generates the corresponding mock observations with the forward simulator, and evaluates the trained network on each of them to obtain the corresponding posteriors. For a well-calibrated posterior, a nominal $p\%$ credible region should contain the true parameter value in $p\%$ of these trials, so plotting the empirical
coverage against the nominal credibility should trace the diagonal $y=x$. Coverage above (below) the diagonal signals posteriors that are too broad (too narrow). \ 

\begin{figure}[ht!]
    \centering
    \includegraphics[width=0.97\textwidth]{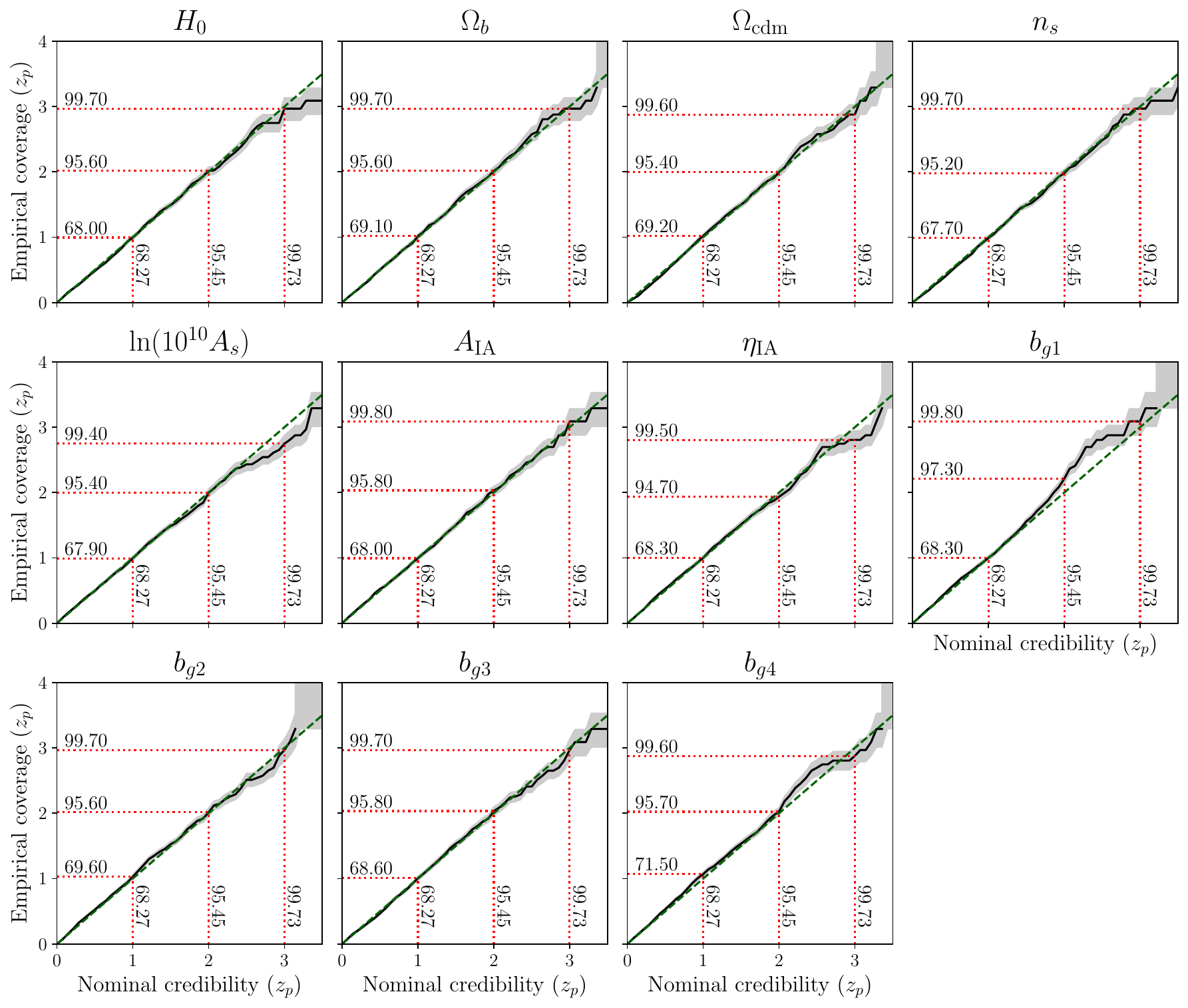}
    \caption{Empirical coverage as a function of nominal credibility for the same parameters as those shown in Fig.~\ref{fig:COSMO_IA_bias}, corresponding to the \threetwopt~analysis with reduced sky fraction. The diagonal indicates perfect calibration.}
    \label{fig:coverage}
\end{figure}

Fig.~\ref{fig:coverage} shows the results of this coverage test for the cosmological, intrinsic-alignment and galaxy bias parameters of the reduced sky-fraction variant, using a batch of 1000 simulations. Rather than working with $p$ directly, we reparametrize it in terms of $z_p$, defined via
\begin{equation}
\frac{p}{100} = \frac{1}{\sqrt{2\pi}} \int_{-z_p}^{z_p} dz\, e^{-z^2/2},
\end{equation}
so that the familiar $(1,2,3)\sigma$ regions correspond to $z_p = (1,2,3)$  with $p = (68.27, 95.45, 99.97)$. This places more emphasis on the posterior tails. We estimate the uncertainty on the empirical coverage arising from the finite number of samples using the Jeffreys interval \citep{Cole:2021gwr}. We find excellent coverage across all parameters, providing further validation of \Merlin. We stress that, while necessary, this test alone is not sufficient to guarantee the perfect calibration of the posteriors; developing more comprehensive validation tools for SBI remains an active area of research \citep{Lemos2023}.  

\end{appendix}

\end{document}